\documentclass[reprint,twocolumn,citeautoscript,superscriptaddress,showpacs,amsmath,amssymb,prb]{revtex4-1}
\usepackage{graphicx}
\usepackage{xcolor}
\usepackage{dcolumn}
\usepackage{bm}
\usepackage{times}
\usepackage[normalem]{ulem}

\begin{document}
\title{Spin polarization of Ru in superconducting Ba(Fe$_{0.795}$Ru$_{0.205}$)$_2$As$_2$ \\studied by x-ray resonant magnetic scattering}
\author{M.~G.~Kim}\email{mgkim@lbl.gov}
\affiliation{Materials Sciences Division, Lawrence Berkeley National Laboratory, Berkeley, California 94720, USA}
\author{J.~Soh}
\affiliation{Ames Laboratory, U.S. DOE and Department of Physics and Astronomy, Iowa State University, Ames, IA 50011, USA}
\author{J.~Lang}
\affiliation{Advanced Photon Source, Argonne National Laboratory, Argonne, Illinois 60439, USA}
\author{M. P. M. Dean}
\affiliation{\mbox{Department~of~Condensed~Matter~Physics~and~Materials~Science, Brookhaven~National~Laboratory,~Upton,~New~York~11973,~USA}}
\author{A.~Thaler}
\affiliation{Ames Laboratory, U.S. DOE and Department of Physics and Astronomy, Iowa State University, Ames, IA 50011, USA}
\author{S.~L.~Bud'ko}
\affiliation{Ames Laboratory, U.S. DOE and Department of Physics and Astronomy, Iowa State University, Ames, IA 50011, USA}
\author{P.~C.~Canfield}
\affiliation{Ames Laboratory, U.S. DOE and Department of Physics and Astronomy, Iowa State University, Ames, IA 50011, USA}
\author{E.~Bourret-Courchesne}
\affiliation{Materials Sciences Division, Lawrence Berkeley National Laboratory, Berkeley, California 94720, USA}
\author{A.~Kreyssig}
\affiliation{Ames Laboratory, U.S. DOE and Department of Physics and Astronomy, Iowa State University, Ames, IA 50011, USA}
\author{A.~I.~Goldman}
\affiliation{Ames Laboratory, U.S. DOE and Department of Physics and Astronomy, Iowa State University, Ames, IA 50011, USA}
\author{R.~J.~Birgeneau}
\affiliation{Materials Sciences Division, Lawrence Berkeley National Laboratory, Berkeley, California 94720, USA}
\affiliation{Department of Physics, University of California, Berkeley, California 94720, USA}
\affiliation{Department of Materials Science and Engineering, University of California, Berkeley, California 94720, USA}

\date{\today}


\begin{abstract}
We have employed the x-ray resonant magnetic scattering (XRMS) technique at the Ru $L_2$ edge of the Ba(Fe$_{1-x}$Ru$_x$)$_2$As$_2$ ($x = 0.205$) superconductor. We show that pronounced resonance enhancements at the Ru $L_2$ edge are observed at the wave vector which is consistent with the antiferromagnetic propagation vector of the Fe in the undoped BaFe$_2$As$_2$. We also demonstrate that the XRMS signals at the Ru $L_2$ edge follow the magnetic ordering of the Fe with a long correlation length, $\xi_{ab} > 2850\pm400$~\AA . Our experimental observation shows that the Ru is spin-polarized in Ba(Fe$_{1-x}$Ru$_x$)$_2$As$_2$ compounds.
\end{abstract}


\maketitle
\section{Introduction}
Superconductivity in the $A$Fe$_2$As$_2$-based ($A$ = Ca, Sr, and Ba) compounds appears as magnetic order is suppressed by substitution on the $A$, Fe or As sites.\cite{Johnston_2010, CandB, PandG, Stewart_2011}  However, the precise role that these substitutions play, particularly for the case of Transition Metal (TM) dopants for Fe, is still a matter of some debate.\cite{Berlijn_12, Fernandes_en_12, kim_cu_2012}  TM substitutions for Fe, including Co,~\cite{sefat_superconductivity_2008, ni_effects_2008} Ni,~\cite{li_superconductivity_2009, canfield_2009} Rh,~\cite{ni_phase_2009, han_2009} Pd,~\cite{ni_phase_2009, han_2009} Ir,~\cite{han_2009} and Pt,~\cite{saha_superconductivity_2010} are generally classified as electron-doping and result in similar phase diagrams for Ba(Fe$_{1-x}$TM$_x$)$_2$As$_2$. For relatively small $x$, both the structural (tetragonal-to-orthorhombic) and the antiferromagnetic (AFM) transition temperatures ($T_S$, $T_N$) are suppressed with $T_S > T_N$, and superconductivity emerges over a small compositional range as doping $x$ increases.\cite{Johnston_2010, CandB, PandG, Stewart_2011, Rotundu_heat_2010, kim_character_2012, sefat_superconductivity_2008, ni_effects_2008, li_superconductivity_2009, canfield_2009, ni_phase_2009, han_2009, saha_superconductivity_2010, sharma_superconductivity_2010, thaler_physical_2010, kim_2011}  In the case of Ba(Fe$_{1-x}$Cu$_x$)$_2$As$_2$, Cu appears to manifest strong impurity scattering effects\cite{fernandes_tc_2012, kim_cu_2012} and superconductivity is not observed, although $T_S$ and $T_N$ are progressively suppressed. Nominal hole-doping through TM substitutions, including Cr \cite{marty_cr_2011} and Mn \cite{kim_mn_2010} also suppresses $T_N$ and $T_S$, but superconductivity is not realized for any level of substitution. In these cases, neutron diffraction measurements~\cite{marty_cr_2011} indicate that G-type AFM order appears at higher Cr concentrations in Ba(Fe$_{1-x}$Cr$_x$)$_2$As$_2$ and recent inelastic neutron scattering measurements\cite{tucker_2012}  on Ba(Fe$_{1-x}$Mn$_x$)$_2$As$_2$ have revealed that G-type spin fluctuations are present in coexistence with static stripe-like AFM order.  The presence of alternative AFM order/fluctuations in these cases may be related to the absence of superconductivity.

Ru substitution in Ba(Fe$_{1-x}$Ru$_x$)$_2$As$_2$ presents a particularly interesting case since it is isovalent with iron and, therefore, would not be expected to contribute additional charge carriers to the system.  Nevertheless, Ru substitution induces superconductivity upon suppression of the stripe-like AFM order albeit at much higher concentrations than other TM elements.\cite{sharma_superconductivity_2010, thaler_physical_2010, kim_2011} The question of whether Ru donates additional charge carriers has been a matter of some debate. Some band structure calculations\cite{sharma_superconductivity_2010} and angle resolved photoemission spectroscopy (ARPES) measurements\cite{brouet_2010, Xu_arpes_2012} (at high temperature in the paramagnetic phase) noted significant differences in the Fermi surface (FS) between the parent BaFe$_2$As$_2$ compound and Ba(Fe$_{0.65}$Ru$_{0.35}$)$_2$As$_2$, concluding that Ru substitution introduces extra electrons, changing the size of electron- and/or hole-pockets. However, other theoretical and experimental studies have found that neither the carrier concentration nor the electronic structure change upon Ru substitution in the closely related oxypnictide compounds.\cite{tropeano_2010, nakamura_2011} Furthermore, low-temperature ARPES investigations\cite{dhaka_2011} found no evidence of changes in the Fermi surface (FS) of Ba(Fe$_{1-x}$Ru$_x$)$_2$As$_2$ over a wide range of Ru substitution, in contrast to the previous ARPES measurements. \cite{brouet_2010} We should note that our x-ray resonant magnetic scattering work, which will be presented in this paper, will not resolve the issue of isovalency of Ru doping in these materials.

Another intriguing issue with Ru substitution is its role in the antiferromagnetism of the system. It has been well established that the suppression of AFM is believed to be a crucial ingredient for superconductivity in the Fe-based superconductors.\cite{Johnston_2010, CandB, PandG, Stewart_2011} Nevertheless, the magnetic nature of the transition metal substitutions themselves has not been the focus of much research in this field.  The transition metals in question, namely, Co, Ni, Pt, Ir, or Ru, carry moments in various other compounds, for example, CoO,~\cite{Rechtin_1972} NiO,~\cite{DeBergevin_1972} UPtGe,\cite{mannix_2000} Sr$_2$IrO$_4$,~\cite{Subramanian_1994} and Ca$_2$RuO$_4$~\cite{Zegkinoglou_2005}. It has also been anticipated, in the Fe-base superconductors, that the transition metal elements may carry moments and affect the magnetism of the Fe in this system. For instance, density functional theory (DFT) calculations have predicted that Co in BaCo$_2$As$_2$ acts as a magnetic impurity and forms a ferromagnetic ground state.\cite{sefat_DFT_2009} However, experiments showed that BaCo$_2$As$_2$ does not order magnetically\cite{sefat_DFT_2009}, leading to the speculation that Co in the Fe-based superconductors might be nonmagnetic. Interestingly, recent x-ray resonant magnetic scattering (XRMS) measurements at the Ir $L_3$ edge for Ba(Fe$_{1-x}$Ir$_x$)$_2$As$_2$ superconductors observed a spin polarization of 5$d$ Ir dopant atoms.\cite{dean_2012}  Although it is not possible to distinguish between a spontaneous ordering and induced ordering, the results imply that Ir is a magnetic dopant element, and show that the ordering of Ir spins follows the same AFM ordering as the Fe.\cite{dean_2012} 

We have employed the XRMS technique at the Ru $L_2$ edge of the Ba(Fe$_{1-x}$Ru$_x$)$_2$As$_2$ ($x = 0.205$) superconductor in order to study the magnetic nature of the Ru dopant element. In this paper we show that pronounced resonance enhancements at the Ru $L_2$ edge are observed at the propagation vector where the AFM ordering of the Fe had been reported previously.\cite{kim_2011} We also demonstrate that, within the experimental error and the constraints of our measurement, the XRMS signals at the Ru $L_2$ edge follow the magnetic ordering of the Fe. Our experimental observation thus shows that the Ru dopant element acts similarly to the Ir, a magnetic dopant element.

\section{Experiment}
Single crystals of Ba(Fe$_{0.795}$Ru$_{0.205}$)$_2$As$_2$ were grown out of a FeAs self-flux using the conventional high temperature solution growth technique described in Ref.~\onlinecite{thaler_physical_2010}.
A single crystal was chosen from a single growth batch and its composition was measured at 10 positions on the sample using wavelength dispersive spectroscopy showing a combined statistical and systematic error on the Ru composition of not greater than 5\%. \cite{thaler_physical_2010} Another crystal from the same batch has been studied by high-resolution x-ray diffraction and elastic neutron diffraction measurements.
From our previous neutron diffraction measurements we found that the structural and antiferromagnetic transitions occur simultaneously and undergo a second-order transition from the high-temperature paramagnetic tetragonal structure to the low-temperature antiferromagnetic orthorhombic structure.\cite{kim_2011} We note that the order (i.e. first- or second-) of the structural/antiferromagnetic transitions in Ba(Fe$_{1-x}$Ru$_x$)$_2$As$_2$ needs to be confirmed by high-resolution diffraction measurements performed with great care. For example, recent high resolution x-ray measurements on the undoped BaFe$_2$As$_2$ compound revealed that the seemingly second-order structural/antiferromagnetic transitions with $T_S = T_N$ are actually a second-order structural transition followed by a first-order magnetic transition with $T_S > T_N$.\cite{Rotundu_heat_2010, kim_character_2012}

For the XRMS measurements, one crystal from the previously studied batch (but not the same piece) with dimensions of 5~$\times$~2~$\times$~0.07~mm$^3$ was selected. The extended surface of the crystal was perpendicular to the \textbf{c} axis.
The measured mosaicity of the crystal was less than 0.02$^{\circ}$ full-width-at-half-maximum (FWHM), attesting to the high quality of the sample. The XRMS experiment was conducted on the beam line 4ID-D at the Advanced Photon Source at Argonne National Laboratory at the Ru $L_2$ edge ($E$~=~2.967~keV, $\lambda$~=~4.183~\AA). The beam path on this beam line was enclosed in a vacuum to minimize the absorption of the x-ray beam by air.
The incident radiation was linearly polarized perpendicular to the vertical scattering plane ($\sigma$ polarized) with a spatial cross section of 0.5~mm (horizontal) $\times$ 0.2~mm (vertical).
In this configuration, dipole resonant magnetic scattering rotates the plane of linear polarization into the scattering plane ($\pi$ polarization).

The sample was mounted at the end of the cold finger of a displex cryogenic refrigerator with the tetragonal ($H$,~$H$,~$L$) plane coincident with the scattering plane.
Here we will generally use the tetragonal notation ($H$,~$H$,~$L$) and, where necessary, employ the orthorhombic notation ($H$,~$K$,~$L$)$_O$ with a subscript ``O".
To minimize the absorption of the x-ray beam at this low energy, we used a single Be dome which resulted in $\sim$ 5\% absolute transmission.
Si(1,~1,~1) was used as a polarization analyzer, providing a scattering angle of 83.6$^{\circ}$, to suppress the charge and fluorescence background by 2 orders of magnitude relative to the XRMS signal.
The scattered x-rays were detected using a SII Vortex silicon drift diode coupled with a Canberra 2025 amplifier. The pulse shaping time was set to 0.5 $\mu$s yielding a detector energy resolution of $\sim$180~eV.
A multichannel analyzer was used to separately monitor the elastically scattered x-rays and either the Ru $L_{\alpha 1}$ or $L_{\beta 1}$ fluorescence signals during scans.

\begin{figure}[b]
\begin{center}
\includegraphics[width=1.0\linewidth]{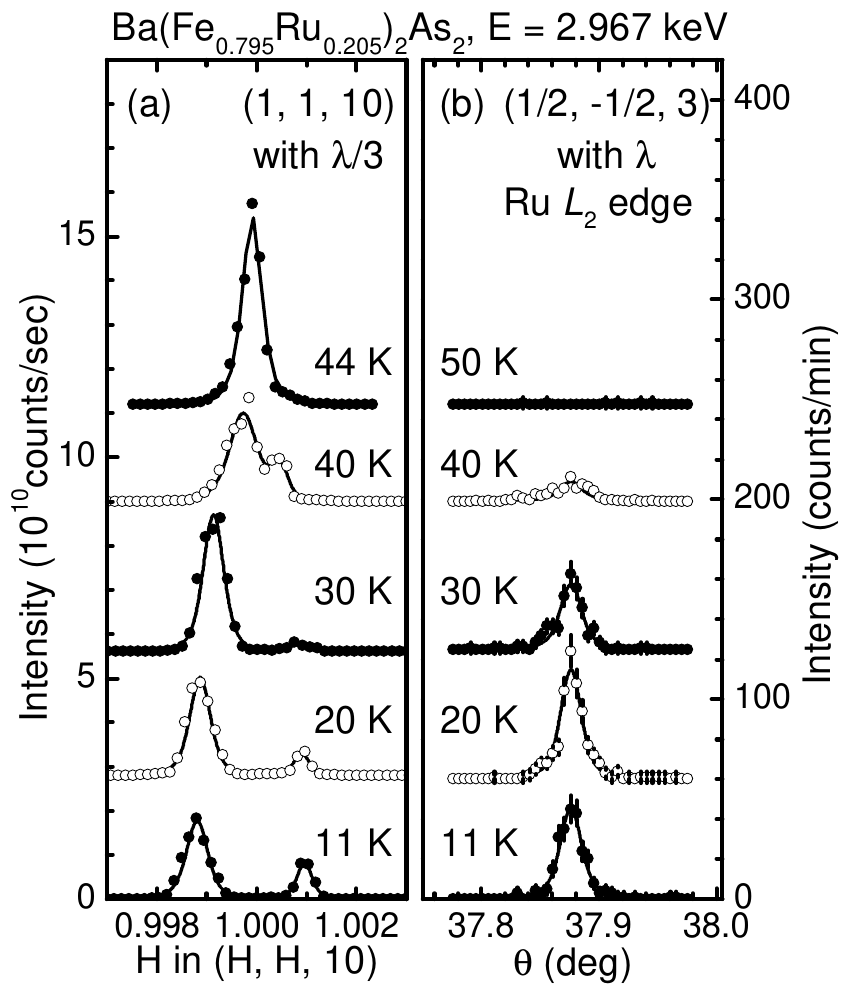}\\
\caption{Temperature evolution of (a) [$H$,~$H$,~0] scan through the (1,~1,~10) Bragg peak measured with off-resonance $E'$ = 8.901~keV ($\lambda'$ = 1.394~\AA) and (b) XRMS signal at the ($\frac{1}{2}$,~$-\frac{1}{2}$,~3) Bragg peak position with the fundamental component ($E$ = 2.967~keV and $\lambda$ = 4.183~\AA) in Ba(Fe$_{0.795}$Ru$_{0.205}$)$_2$As$_2$. The data are shown with arbitrary offsets. The lines present the fitted curves using a Lorentzian-squared line shape.}
\label{fig1}
\end{center}
\end{figure}

Due to the long wavelength of the x-ray at the Ru $L_2$ edge, accessible Bragg reflections were limited. Therefore, whereas the XRMS measurements were performed at an off-specular $(\frac{1}{2},~-\frac{1}{2},~3)$ Bragg peak position with the fundamental component ($E$ and $\lambda$), the measurements at charge peaks with large scattering angles [e.g. (1,~1,~10) reflection] were conducted using the third harmonic component ($E'$~=~3$\times E$~=~8.901~keV, $\lambda'$ = $\frac{\lambda}{3}$ = 1.394~\AA).
Note that the third harmonic component was obtained with no change in the experimental configuration. The data were obtained as a function of temperature between 50~K and 11~K, the base temperature of the refrigerator.

\section{Results and discussion}

Figure~\ref{fig1} (a) displays the temperature evolution of the (1,~1,~10) Bragg peak for Ba(Fe$_{0.795}$Ru$_{0.205}$)$_2$As$_2$.
The incident x-ray was tuned to the Ru $L_2$ edge, but the (1,~1,~10) Bragg peak was measured using the third harmonic component as described in the previous section.
A sharp single (1,~1,~10) Bragg peak of the tetragonal phase at $T$ = 44~K splits continuously into two peaks [(2,~0,~10)$_O$ and (0,~2,~10)$_O$] of the orthorhombic phase for temperatures below $T_S$ = 43$\pm$1~K.
As temperature decreases further, the orthorhombic distortion $\delta$ = $\frac{a-b}{a+b}$ increases and reaches $\sim$11$\times$10$^{-4}$ at $T$ = 11~K.

Above $T$ = 44~K no XRMS signal is observed at Q~=~($\frac{1}{2}$,~$-\frac{1}{2}$,~3), but as the temperature is lowered, a clear resonant enhancement is observed in the $\sigma$-$\pi$ scattering channel and the XRMS signals increase progressively [Fig.~\ref{fig1}~(b)]. The XRMS signal at the Ru $L_2$ edge at low temperature corresponds to the dipole resonant process, exciting 2$p$ core electrons into the 4$d$ valence band. The propagation vector at which the XRMS signal is observed is identical to the antiferromagnetic propagation vector Q$_{\textrm{AFM}}$ for BaFe$_2$As$_2$ compounds indicating that the Ru spin polarization is the same as observed for the Fe, an AFM alignment of the moments along the orthorhombic $\textbf{a}$ and $\textbf{c}$ axes and FM alignment along the $\textbf{b}$ axis. Using the correlation length defined as $\xi = 1/\omega$, with $\omega$ as the half-width-at-half-maximum of the diffraction peak in the inverse length scale, we find the magnetic correlation length in the $ab$ plane, $\xi_{{ab}} >$ $2850\pm400$ \AA~which indicates that spins on the Ru site are well correlated.

\begin{figure}[t]
\begin{center}
\includegraphics[clip, width=1.0\linewidth]{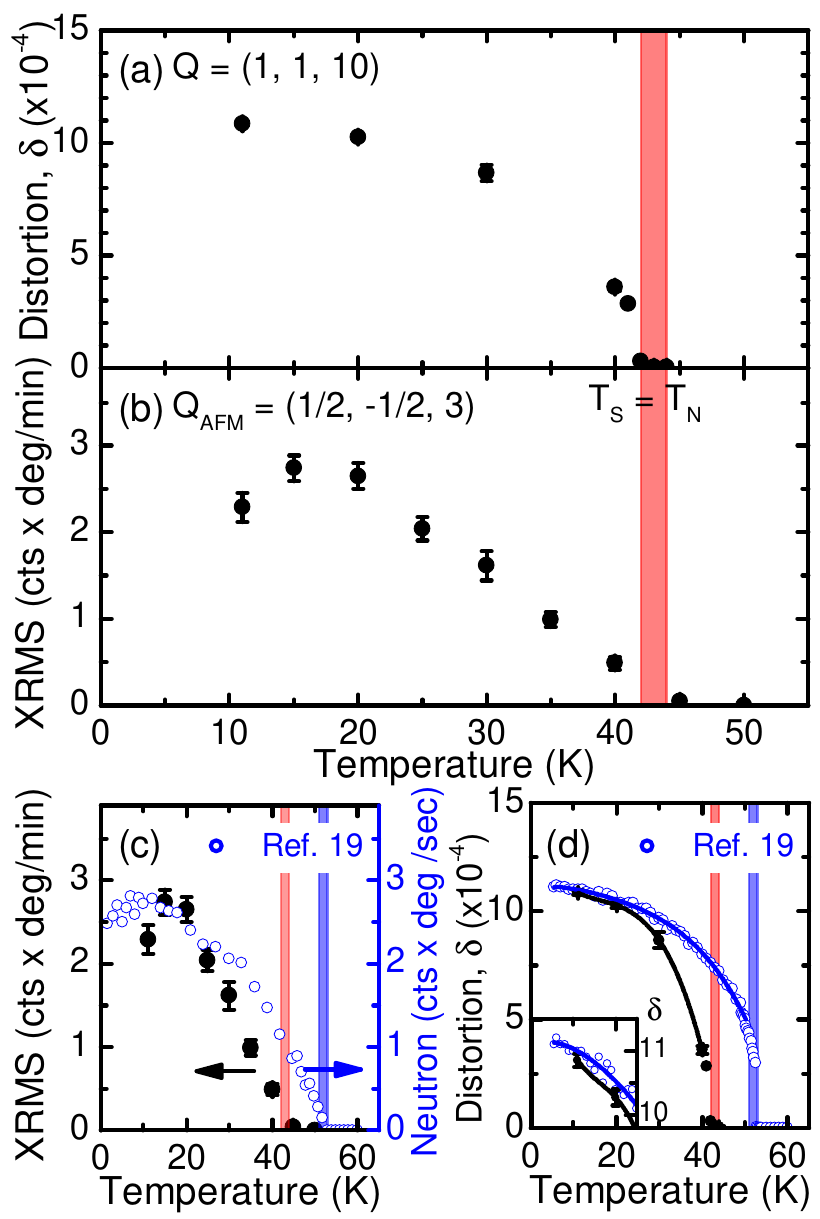}\\
\caption{(Color online) (a) Temperature dependence of the orthorhombic distortion $\delta$ = $\frac{a-b}{a+b}$ upon warming determined from fits to the (1,~1,~10) Bragg peak.  (b) The evolution of the integrated intensities of the ($\frac{1}{2}$,~$-\frac{1}{2}$,~3) XRMS peak as a function of temperature during warming. The red bars indicate the structural/AFM transition temperatures, indicating that $T_S = T_N = 43\pm1 $ K. The temperature dependent XRMS signals and the orthorhombic distortion in the current work are compared with (c) the AFM ordering of the Fe moment and (d) the orthorhombic distortion observed in Ref.~\onlinecite{kim_2011}, respectively. Transition temperatures determined in Ref.~\onlinecite{kim_2011} are marked with blue bars. The discrepancy in transition temperatures is likely due to the sample heating by the strong incident x-ray beam as described in the text.}
\label{fig2}
\end{center}
\end{figure}

In Figures.~\ref{fig2} (a) and (b) we show the orthorhombic distortion, $\delta$ = $\frac{a-b}{a+b}$, of the (1,~1,~10) Bragg peak and the integrated intensity of the ($\frac{1}{2}$,~$-\frac{1}{2}$,~3) XRMS peak as functions of temperature.  The orthorhombic distortion and the evolution of the XRMS signal appear at a very close temperature as indicated by the red bar in Figs.~\ref{fig2} (a) and (b). However, a comparison to the previous measurements ($T_S$ = $T_N$ = 52$\pm$1~K)~\cite{kim_2011} shows that the structural and antiferromagnetic transitions in the current work appear at about 9 K lower as shown in Figs.~\ref{fig2} (c) and (d). Despite the discrepancy in observed transition temperatures, we can conclude that the  XRMS signals from the Ru $L_2$ edge appear at the AFM transition temperature of the Fe because it is known that, within experimental error, the structural and AFM transitions of the Fe are concomitant in temperature in Ru substituted BaFe$_2$As$_2$ compounds.\cite{thaler_physical_2010,kim_2011}

We attribute the offset in temperature to the large absorption, and consequent sample heating, of the long wavelength incident x-rays by the sample, compounded by the use of only a single Be dome as a heat shield and the absence of exchange gas, necessary to minimize x-ray absorption. From our experience in measuring the orthorhombic distortion in various compounds\cite{kim_2011, kim_mn_2010,kim_character_2012,dean_2012,kim_parent_2010}, we have found that the degree of orthorhombic distortion exhibits almost identical values in compounds possessing the same substitution element and composition levels. Therefore, the values of the orthorhombic distortion at low temperature in two measurements indicate that the sample studied by XRMS is very similar to the previous sample studied by neutron diffraction [Fig.~\ref{fig2} (d)]. The sample heating effect can be also seen by comparing the size of the distortions at given temperatures. The orthorhombic distortion at the base temperature $T$ = 11~K, $\delta \sim11\times10^{-4}$, in our current measurement is closer to the value measured at $\sim$17~K in our previous laboratory measurements, which gives about a 6~K temperature difference, and $\delta \sim4\times10^{-4}$ at 40~K (current work) and 51~K (previous work) shows an 11~K difference in transition temperature [Fig.~\ref{fig2} (d)]. The range of temperature differences (6~K$-$11~K) can be understood by the different cooling power of the refrigerator, which performs stronger cooling at lower temperature resulting in less sample heating. We conclude that the offset in temperature between the present XRMS measurements and our previous neutron diffraction study is due to sample heating. We further note that a sudden drop of the integrated intensity of the ($\frac{1}{2}$,~$-\frac{1}{2}$,~3) XRMS peak at $T$ = 11~K is likely an artifact and not an indication of a suppression of the Ru spin ordering below $T_c \approx$  13~K because no such behavior was present when the signals were measured while sitting on top of the peak (not shown).

The observed XRMS signals at ($\frac{1}{2}$,~$-\frac{1}{2}$,~3) [Fig.~\ref{fig1} (b)] together with its temperature dependence  [Fig.~\ref{fig2} (b)] demonstrate that the Ru dopant atoms are spin-polarized and the spin polarization follows the AFM ordering of the Fe in Ba(Fe$_{0.795}$Ru$_{0.205}$)$_2$As$_2$. However, as discussed in Ref.~\onlinecite{dean_2012}, we can not conclude whether the spin polarization of the Ru is induced by the local field from the Fe neighbors or via other indirect interactions between the Ru and Fe states. 

\begin{figure}
\begin{center}
\includegraphics[clip, width=0.95\linewidth]{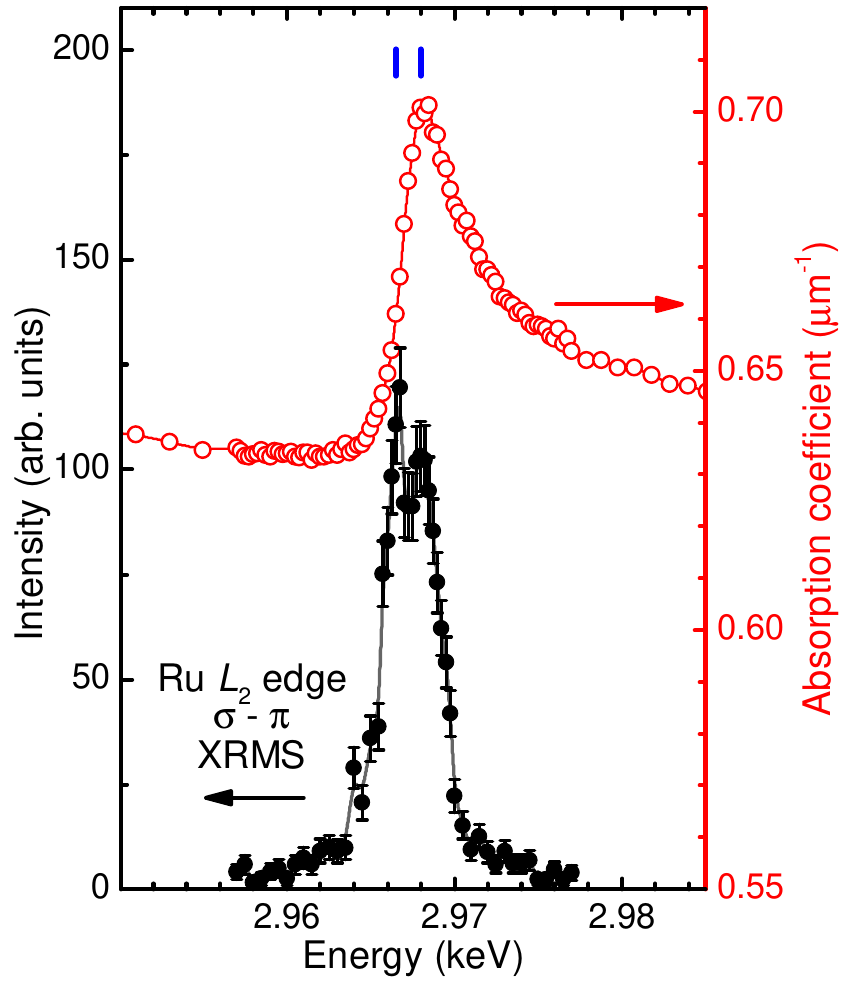}\\
\caption{(Color online) Energy scan through the ($\frac{1}{2}$,~$-\frac{1}{2}$,~3) XRMS peak (filled circles) and the energy dependent absorption coefficient (open circles) calculated from the fluorescence spectrum measured at 45$^{\circ}$ scattering angle around the Ru $L_2$ edge. The energy scan is corrected for absorption. Blue bars indicate positions of two resonant peaks. Lines are guides to eyes.} \label{fig3}
\end{center}
\end{figure}

Figure~\ref{fig3} shows an absorption corrected energy scan around the Ru $L_2$ edge ($E$ = 2.967~keV) in the $\sigma$-$\pi$ scattering geometry at a constant Q = ($\frac{1}{2}$,~$-\frac{1}{2}$,~3) at $T$ = 11~K (filled circles) and the energy dependence of the absorption coefficient (open circles) as calculated from the fluorescence spectrum as described in Ref.~\onlinecite{Bruckel_2001}. We notice that the resonant energy spectrum consists of two well-defined peaks: a peak at $E$ = 2.9665~keV where the inflection point is present in the fluorescence spectrum, and a second peak at 1.5~eV higher energy ($E$ = 2.968~keV) where the fluorescence is maximum.

Two peaks in the XRMS energy scan around the Ru $L_{2}$ edge have been observed in Ruthenates such as Ca$_2$RuO$_4$ and Ca$_3$Ru$_2$O$_7$.\cite{Zegkinoglou_2005, Bohnenbuck_2008}
In Ca$_2$RuO$_4$, both orbital and magnetic order are present and the orbital ordering emerges at higher temperature than the magnetic ordering temperature;
the two resonant peaks at the Ru $L$ edges are temperature dependent, changing both the spectral weights and positions, because of the different resonant responses from the orbital and the magnetic order.\cite{Zegkinoglou_2005} Ca$_3$Ru$_2$O$_7$ has been also claimed to display an orbital ordering, but that has not yet been confirmed.\cite{Bohnenbuck_2008}

In a similar vein, it is possible that Ru orbital ordering (either spontaneous or induced polarization, and likely anti-ferro) exists in Ba(Fe$_{0.795}$Ru$_{0.205}$)$_2$As$_2$.
The two peaks in the energy spectrum for Ba(Fe$_{0.795}$Ru$_{0.205}$)$_2$As$_2$ may reflect  resonant transitions into different Ru 4$d$ orbitals (e.g. 4$d~t_{2g}$ and 4$d~e_g$ orbitals), and these orbitals may contribute differently to the resonance process.
However, we can not exclude the possibility that the observed two peaks may be the features common in resonant scattering of $d$-electron elements in the FeAs-based superconductors. For example, the energy scan around the Fe $K$ edge for the parent BaFe$_2$As$_2$ exhibits a sharp peak close to the absorption threshold and broad features extending up to $\sim$20~eV~\cite{kim_parent_2010} although the two peaks in Ba(Fe$_{0.795}$Ru$_{0.205}$)$_2$As$_2$ are much closer and appear in a narrower energy range than the features in the parent compound.

\section{summary}

In summary, we have studied the spin polarization of the Ru 4$d$ dopant elements in the Ba(Fe$_{0.795}$Ru$_{0.205}$)$_2$As$_2$ compound. A sample of Ba(Fe$_{0.795}$Ru$_{0.205}$)$_2$As$_2$ presents a structural phase transition at $T_S$ = 43$\pm$1~K. The resonance enhancement at the Ru $L_2$ edge appears at $\approx T_S$ at Q$_\mathrm{AFM}$ = (1/2, -1/2, 3), consistent with the AFM propagation vector of the Fe order. Despite the fact that the observed transition temperatures are lower than previous reports on the same Ru composition, the concurrent appearance of the orthorhombic splitting and the XRMS signal indicates that the spin polarization of the Ru dopant element emerges at a temperature ($T_S$) where the AFM order of the Fe also emerges. We also show that the spins on the Ru dopant atoms are correlated over $>$ 700 unit cells in the \textbf{ab} plane. Thus, the Ru is a magnetic dopant element. From the observation of two well-defined peaks in the resonant energy spectrum around the Ru $L_2$ edge, we propose that the Ru 4$d$ orbitals may be polarized contributing to different resonant processes. Further theoretical and experimental studies would be beneficial to understand the observed energy spectrum  in Ru substituted BaFe$_2$As$_2$ superconductors.

\begin{acknowledgments}
We would like to thank Y. Choi and J. W. Kim for valuable discussions and help in the experiment. This work was supported by the U.S. Department of Energy (DOE), Office of Basic Energy Sciences, Materials Sciences and Engineering Division, under Contract No. DE-AC02-05CH11231.
The work at the Ames Laboratory was supported by the Department of Energy-Basic Energy Sciences under Contract No. DE-AC02-07CH11358. The work at Brookhaven is supported in part by the US DOE under Contract No. DE-AC02-98CH10886 and in part by the Center for Emergent Superconductivity, an Energy Frontier Research Center funded by the US DOE, Office of Basic Energy Sciences. Use of the Advanced Photon Source, an Office of Science User Facility operated for the U.S. DOE Office of Science by Argonne National Laboratory, was supported by the U.S. DOE under Contract No. DE-AC02-06CH11357.
\end{acknowledgments}

\bibliographystyle{apsrev}
\bibliography{XRMS_ru}

\end{document}